# Application of Variance-Based Sensitivity Analysis to a Large System Dynamics Model

Daniel Inman Laura J. Vimmerstedt Brian Bush Dana Stright
Steve Peterson

**Acknowledgments**

This work was supported by the U.S. Department of Energy under Contract No. DE-AC36-08-GO28308 with the National Renewable Energy Laboratory, 15013 Denver West Parkway, Golden, CO 80401-3305.

**Abstract**

Variance-based sensitivity methods can provide insights into large computational models. We present a novel application of sensitivity analysis to the Biomass Scenario Model (BSM) a large and complex system dynamics model of the developing biofuels industry in the United States. We apply a two-stage sensitivity approach consisting of an initial sensitivity screening, followed by a variance decomposition approach. Identifying key system levers and quantifying their strength is not straightforward in complex system dynamics models that have numerous feedbacks and nonlinear results. Variance-based sensitivity analysis (VBSA) offers a systematic, global approach to assessing system dynamics models because it addresses nonlinear responses and interactive effects. Especially when a large model's size makes manual exploration of the input space difficult and time-consuming, the approach can help to provide a comprehensive understanding of interactions that drive model behaviors.

# 1 Introduction

Large computational models, such as the Biomass Scenario Model (BSM), are challenging to analyze. Such models are rarely examined through rigorous statistically-designed studies and analyses. Sensitivity analysis is one approach that can be used to quantitatively assess a model's input factors to determine which ones are the most influential with regard to a specific model output metric. This information is valuable for users of large models because (a) it provides a quantitative means for vetting the model, and (b) it can lead to insights about the system(s) being modeled that would be difficult to gain using a scenario-driven approach to exercising the model.

The objective of this study is to apply variance-based sensitivity analysis to the BSM to gain nontrivial system insights. Specifically, we use sensitivity analysis to identify the most influential biofuels incentives and interactions among incentives. Many of the inputs are uncertain or unknown, and it is unclear, *a priori*, which incentives or combinations of incentives will be most effective for promoting biofuels production.

This paper is organized as follows: Section 2 introduces the methods used in this study, including details related to the BSM itself and variance-based sensitivity analytical methods. Our main results are presented in Section 3. Finally, we provide a discussion in Section 4.

# 2 Methods

We use a two-stage approach to sensitivity analysis in this study. First, we identify a set of model factors that are of interest. Because variance-based sensitivity analysis (VBSA) is computationally intense, we use elementary effects analysis to select a smaller subset of model factors that appear to have significant influence on the model's output. Second, we perform VBSA on this smaller subset of factors to determine the total effects, first-order, and second-order sensitivity indices.

## 2.1 Biomass Scenario Model

The BSM is a large and complex system dynamics model of the biofuels supply chain in the United States, developed in STELLA, a commercial system dynamics modeling software package. The BSM uses a system of coupled ordinary differential equations to model the trajectory of the system's development through time. The model has numerous nonlinear feedbacks and interactions represented in it's structure. For a more detailed description of the BSM see ([16]).

## 2.2 Model Factors

The factors and levels selected were intended to focus on policy incentives designed to spur industrial growth ([10]). Table 1 lists the factors varied in this study. The factors explored in this study are not inclusive of all possible factors and combinations in the BSM. Tables 2 and 3 show the BSM inputs that represent the incentives that were examined.

Table 1: Biorefinery attributes included in the initial 140 factors used in elementary effects screening

| Pathway Attribute | Categories* | | Number of Attributes |
|---|---|---|---|
| Feedstock and Product | Cellulose to ethanol | Cellulose to hydrocarbons | 2 |
| Conversion Technology | Biochemical (BC) | Catalytic Conversion of Sugars to Hydrocarbons (CCSH) | 2 + 5 |
| | Thermochemical (TC) | Fast Pyrolysis (FP) | |
| | | Fermentation of Sugars to Hydrocarbons (F) | |
| | | Fischer-Tropsch (FT) | |
| | | Indirect Liquefaction to High-Octane Gasoline (MTG) | |
| Scale | Pioneer Commercial | | 2 |
| | Full Commercial | | |

This table shows some of the pathway attributes that BSM uses to characterize each biorefinery, including feedstock and product, conversion technology, and scale. The exploration of incentive values does not imply a judgment that these ranges are feasible, likely, or recommended. The two feedstock and product categories included in the study are listed here: cellulose to ethanol and cellulose to hydrocarbons (*). Within each of these categories, conversion technology attributes are listed: two for cellulose to ethanol and five for cellulose to hydrocarbons. Each biorefinery is represented at two scales of production: pioneer commercial and full commercial. These attributes combined with the incentives in Tables 2 and 3 describe the initial 140 factors; the sum of the combinations of biorefinery attributes with incentive attributes is 140.

*Categories listed are not comprehensive of all pathways in the BSM, as described in text and references.

Table 2: Incentives included in the initial 140 factors used in elementary effects screening: fixed capital investment grants and price support.

| Subsidy Type | Description | Limits | Min | Max | Base | Paths | P and C or Both |
|---|---|---|---|---|---|---|---|
| Fixed Capital Investment | Capital cost subsidy (%) | | 0 | 1 | 0 | EtOH & HC | 2 |
| | | Start year | 2011 | 2015 | | 2 + 5 | 2 |
| | | Duration (yr) | 0 | 20 | 0 | 2 + 5 | 2 |
| | Subsidy during start-up (%) | | 0 | 1 | 0 | EtOH & HC | 2 |
| | | Volumetric termination threshold (billion gal) | 0 | 5 | 0 | EtOH & HC | 2 |
| Price | Price subsidy ($/gal) | | 0 | 5 | 0.46<br>1.01 | EtOH<br>HC | Both |
| | | Start year | 2011 | 2015 | | 2 + 5 | Both |
| | | Duration (yr) | 0 | 20 | 0 | 2 + 5 | Both |
| | Subsidy during start-up ($/gal) | | 0 | 5 | 0 | EtOH & HC | Both |
| | | Volumetric termination threshold (billion gal) | 0 | 5 | 0 | EtOH & HC | Both |

This table shows the incentives explored in this study. "Subsidy type" characterizes the application of the subsidy. "Description" provides detail on when the subsidy applies: during a defined time period or during a start-up period limited by production volume. "Limits" describe these temporal and volumetric limits. "Min" and "Max" are the ranges of each factor used in the sensitivity analysis design. The initial year in the model is 2011. "Base" shows the values used in the base case (see also [27]). "Paths" indicate how this factor varies according to the pathway in Table 1: either at the level of "Feedstock and Product" (EtOH [cellulose to ethanol] or HC [cellulose to hydrocarbons]) or with "Conversion Technology" (2 + 5). "P and C or Both" indicates how this factor varies according to the scale in Table 1: either at the level of Pioneer Commercial (P), Full Commercial (C), or Both. This table shows the incentive characteristics that combine with the biorefinery attributes in Table 1 to describe the initial 140 factors; the sum of the combinations of biorefinery attributes with incentive attributes is 140.

Table 3: Incentives included in the initial 140 factors used in elementary effects screening: loan guarantees and feedstock price support.

| Subsidy Type | Description | Limits | Min | Max | Base | Paths | P and C or Both |
|---|---|---|---|---|---|---|---|
| | Loan guarantee (%) | | 0 | 1 | 0 | EtOH & HC | 2 |
| Loan | | Start year | 2011 | 2015 | | 2 + 5 | 2 |
| | | Duration (yr) | 0 | 20 | 0 | 2 + 5 | 2 |
| | Subsidy during start-up (%) | | 0 | 1 | 0 | EtOH & HC | 2 |
| | | Volumetric termination threshold (billion gal) | 0 | 5 | 0 | EtOH & HC | 2 |
| | Feedstock subsidy ($/t) | | 0 | 1 | 0 | EtOH & HC | 2 |
| Feedstock | | Start year | 2011 | 2015 | | 2 + 5 | 2 |
| | | Duration (yr) | 2011 | 2015 | | 2 + 5 | 2 |
| | Subsidy during start-up ($/t) | | 0 | 1 | 0 | EtOH & HC | 2 |
| | | Volumetric termination threshold (billion gal) | 0 | 5 | 0 | EtOH & HC | 2 |

This table shows the incentives explored in this study. “Subsidy type” characterizes the application of the subsidy. “Description” provides detail on when the subsidy applies: during a defined time period or during a start-up period limited by production volume. “Limits” describe these temporal and volumetric limits. “Min” and “Max” are the ranges of each factor used in the sensitivity analysis design. The initial year in the model is 2011. “Base” shows the values used in the base case (see also [27]). “Paths” indicate how this factor varies according to the pathway in Table 1: either at the level of “Feedstock and Product” (EtOH [cellulose to ethanol] or HC [cellulose to hydrocarbons]) or with “Conversion Technology” (2 + 5). “P and C or Both” indicates how this factor varies according to the scale in Table 1: either at the level of Pioneer Commercial (P), Full Commercial (C), or Both. This table shows the incentive characteristics that combine with the biorefinery attributes in Table 1 to describe the initial 140 factors; the sum of the combinations of biorefinery attributes with incentive attributes is 140.

### 2.3 Model Output Metrics

For both the elementary effects and VBSA (discussed below), we use six metrics of biofuels production by fuel classification and timing: maximum total biofuels production prior to 2035, maximum total biofuels production prior to 2051, maximum cellulose-to-hydrocarbons production prior to 2035, maximum cellulose-to-hydrocarbons production prior to 2051, maximum cellulose-to-ethanol production prior to 2035, and maximum cellulose-to-ethanol production prior to 2051. Maxima were used, rather than production in a specific year, because of oscillatory behavior in the time series of simulated fuel production.

### 2.4 Elementary Effects Screening

The elementary effects method is a quick and simple approach to screening a large number of model factors for their influence on model outputs; *e.g.,* [18] and [15]. Because of the large number of BSM inputs, a screening method is helpful to reduce computational requirements. As described below, we sampled the targeted input space, simulated results, and calculated elementary effects sensitivity measures to select the most influential factors from among the 140 initial factors.

Each of the 140 initial factors has a range, such that the input space is large. To sample this input space efficiently, Morris's one-at-a-time (OAT) sampling method was used to create a study design ([15]). This sampling method relies on dividing the input range of each of the $k$ factors into $p$ different levels, selecting unique trajectories within the $p \times k$ input space, and simulating the model results for each trajectory. For this study, we created a study design using the $k = 140$ factors described above, $r = 25$ replications or trajectories, for a total of $r(k+1) =$ 3,525 simulations. We used R statistical programming language ([17]) to implement this study design (see online supplemental material for R code).

Using the results of these 3,525 simulations, we calculated elementary effects sensitivity measures for the influence on model output of each of the 140 input factors over their selected ranges. The sensitivity measures of the elementary effects of each factor on model output were average elementary effects $\mu$ (Eq. 1); the average absolute value of the elementary effects, $\mu^*$ (Eq. 2); and the variance of the elementary effects, $\sigma^2$ (Eq. 3) ([18], pp. 116–117). [20] proposed $\mu^*$ as a more robust measure of a factor's influence because it avoids Type II errors that might result from factors when low values of $\mu$ occur because of summing negative and positive elementary effects. The strength of each factor's interaction with other factors is indicated by the value of $\sigma^2$ ([20]). In this study, we also calculated the standard error of the mean for $\mu^*$ (the square root of Eq. 3). The standard error of $\mu^*$ was used as a guide to determine which factors' $\mu^*$ values were potentially influential; factors with larger standard error are more likely to be influential. This test is nondirectional, meaning that the factors selected could have either a negative or positive impact on the metric (biofuels production).

Factors that were influential for all fuel production classifications were included, but some factors were included because of their influence on cellulose to hydrocarbons alone. We selected a total of 20 potentially influential factors to analyze in the VBSA.

**Average elementary effects:**

$$\mu_i = \frac{1}{r} \sum_{j=1}^{r} EE_i^j \tag{1}$$

**Average absolute value of the elementary effects:**

$$\mu_i^* = \frac{1}{r}\sum_{j=1}^{r}|EE_i^j| \tag{2}$$

**Variance of the elementary effects:**

$$\sigma_i^2 = \frac{1}{r-1}\sum_{j=1}^{r}(EE_i^j - \mu)^2 \tag{3}$$

where

$EE$ is the metric for elementary effect as defined in [15].

$i$ designates each of the $k$ factors.

$r$ is the number trajectories that sample each of the $k$ factors at $p$ different values within their range.

## 2.5 Variance-Based Sensitivity Analysis

From the elementary effects screening, we selected 20 factors that were considered to be influential based on their standard errors (Figure 4) and we used VBSA to estimate their effects. We used a modified version of the VBSA methodology of [21].

We used Sobol quasi-random sequences to generate a sampling hypercube ([23], [18]). A base sample size of $N = 4000$ was selected. In addition to the 20 potentially influential factors, we included three statistical control factors that did not meet the elementary effects screening criteria, as described below. This resulted in a study design based on $N = 4000$ sample size for the $k = 23$ factors, resulting in a total of $N(2k + 2) =$ 192,000 simulations.

We calculated first-order, second-order, and total effects sensitivity indices to determine the influence of and interactions among the selected factors. First-order effects show relative influence on model variance of each factor alone. Second-order effects show relative influence on model variance attributed to each pair of factors that are not explained by their first-order effects. Total effects reflect the contribution of a factor, alone and in combination with other model factors, to the variance in the model's output. The total effects index is often used to determine which factors can be held constant without significantly affecting model variance. Large differences between total effects and first-order effects suggest that the factor is highly interactive.

Equations for model output (4), first- order (5), second-order (6), and total effects indices (7) are shown below. The equations for the indices use partial variances $V_i$ and $V_{ij}$, and the variance decomposition $V_{X_i}$ is the variance-based first-order effect for factor $X_i$ on $Y$, or $V_{X_i}[E_{X_{\sim i}}(Y|X_i)]$.

$$Y = f(X_1, X_2, ...X_k) \tag{4}$$

$$S_i = \frac{V_i}{V(Y)} \tag{5}$$

$$S_{ij} = \frac{V_{ij}}{V(Y)} \tag{6}$$

$$S_{Ti} = \frac{E_{X_{\sim i}}[V_{X_i}(Y|X_{\sim i})]}{V(Y)} = 1 - \frac{V_{X_{\sim i}}[E_{X_i}(Y|X_{\sim i})]}{V(Y)} \tag{7}$$

First-order, second-order, and total effects are reported in the Results section.

The uncertainty estimation used in this study is quantitative and does not account for uncertainty arising from the model and from data quality. We used bootstrapping to provide estimates of the 95% confidence interval for each of the variance-based sensitivity measures ($S_i$, $S_{ij}$, $S_{Ti}$) ([8]).

# 3 Results

Our results include time series of biofuels production for the 3,525 simulations performed in the elementary effects study; time series and frequency distributions of biofuels production for the 192,000 simulations performed in the VBSA study; and VBSA-based estimates of first-order, second-order, and total effects sensitivity indices and their uncertainties. Results are contingent upon the model formulation and input settings, and they should be interpreted with caution.

## 3.1 Model Simulations: Biofuels Production

Figures 1 and 2 display biofuel production throughout the study period for all simulations. The primary difference between the two figures is visible in the top panel, where for the elementary effects screening (Figure 1) we included higher levels of certain factors, which led to growth levels that outstripped the market and thus caused the industry to crash in some instances.

The development of the cellulose-to-ethanol industry occurs earlier and requires less incentive than the cellulose-to-hydrocarbons industry. Cellulose-to-ethanol production rarely exceeds 15 billion gallons per year. In figures 1 and 2, this can be observed in the cellulose-to-ethanol panels, where production rarely exceeds approximately 15 billion gallons. This implicit limit can trigger declines or oscillations in ethanol production, as shown in the figures. If cellulose-to-hydrocarbons production grows, it tends to increase or stabilize in later years (see figures 1 and 2), and it does not show declines or oscillations because it has no implicit limit.

Figure 3 shows the frequency distribution of maximum fuel production volume among production volume bin sizes starting at no production and extending to the highest production volume simulated. Each panel represents one metric assessed. One salient feature is the large number in the smallest production bin, which indicates that the cellulose-to-hydrocarbons industry does not develop in the majority of the model simulations. The frequency distribution shows that the production volume is sensitive within the ranges of input values that were examined, as necessary for meaningful results and intended in the analytic design.

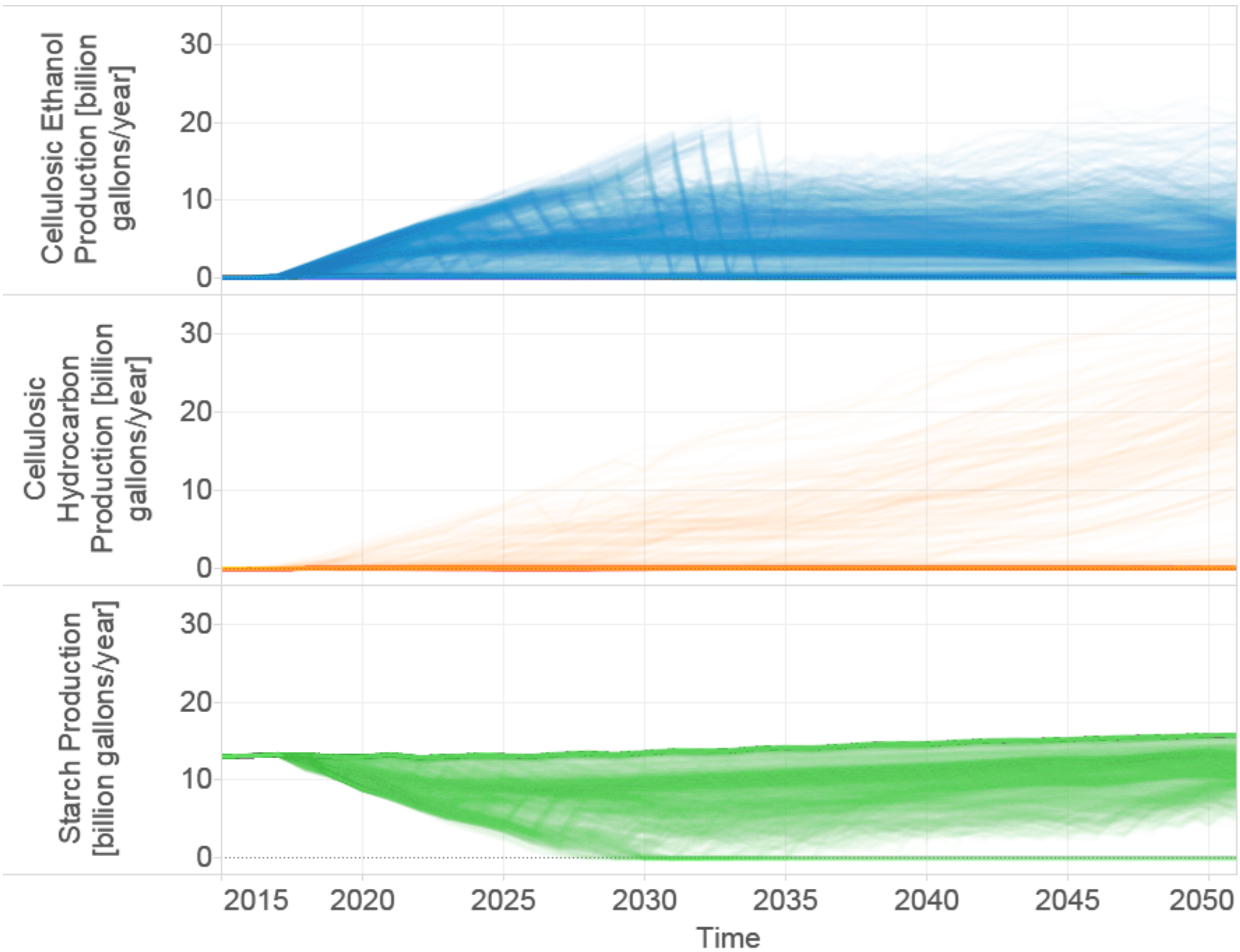

Figure 1: Cellulose-to-ethanol, cellulose-to-hydrocarbons, and starch-to-ethanol production results are shows for elementary effects screening simulations. Each line represents one of the 3525 simulations. The effects of the blend wall (i.e. the blending of ethanol with gasoline at 10 percent by volume) are evident in the paucity of simulations that exceed approximately 15 billion gallons of ethanol production, and in the crashes in cellulose-to-ethanol production in some simulations.

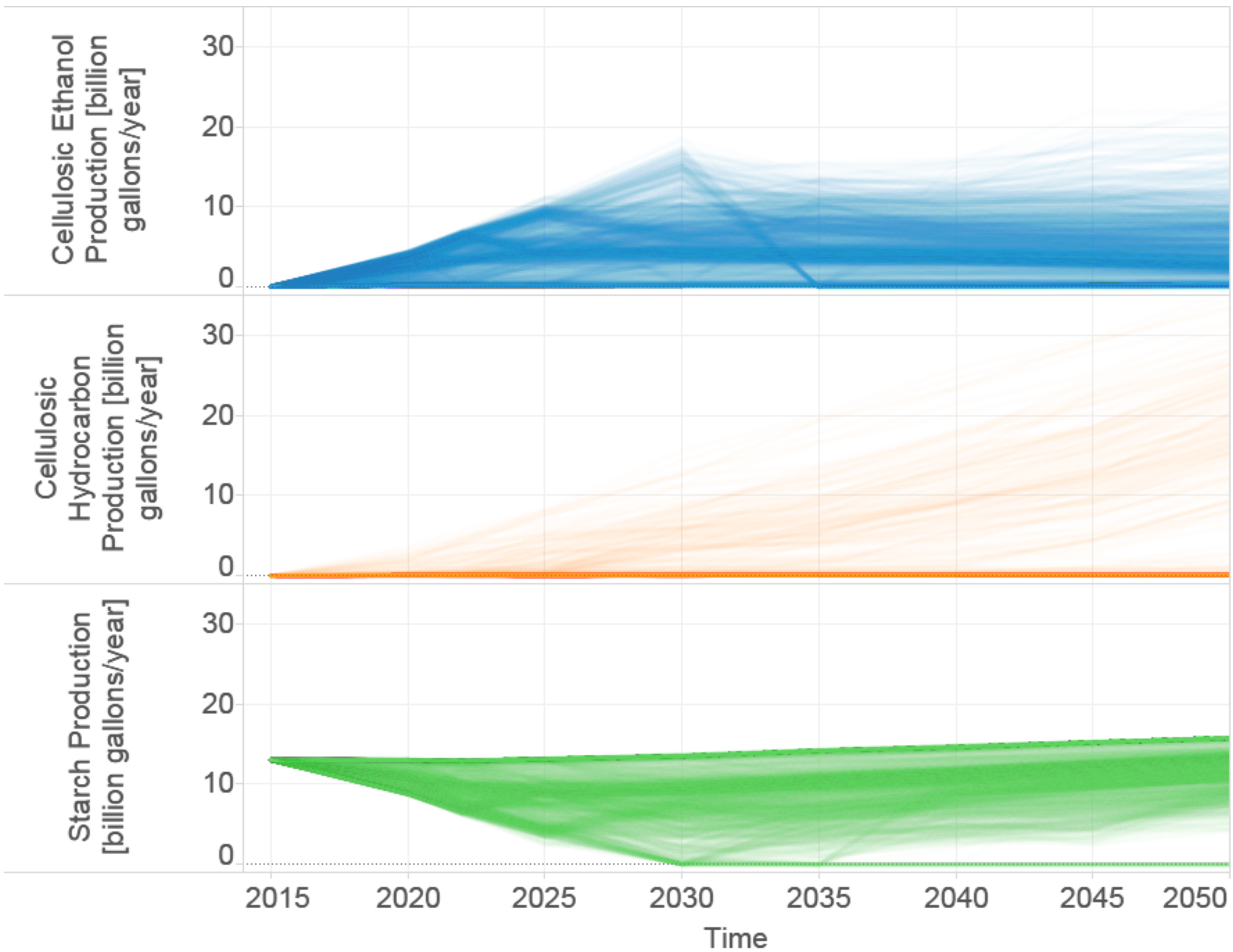

Figure 2: Cellulose-to-ethanol, cellulose-to-hydrocarbons, and starch-to-ethanol production results are shown for all model simulations. Each line represents one of the 192,000 simulations.

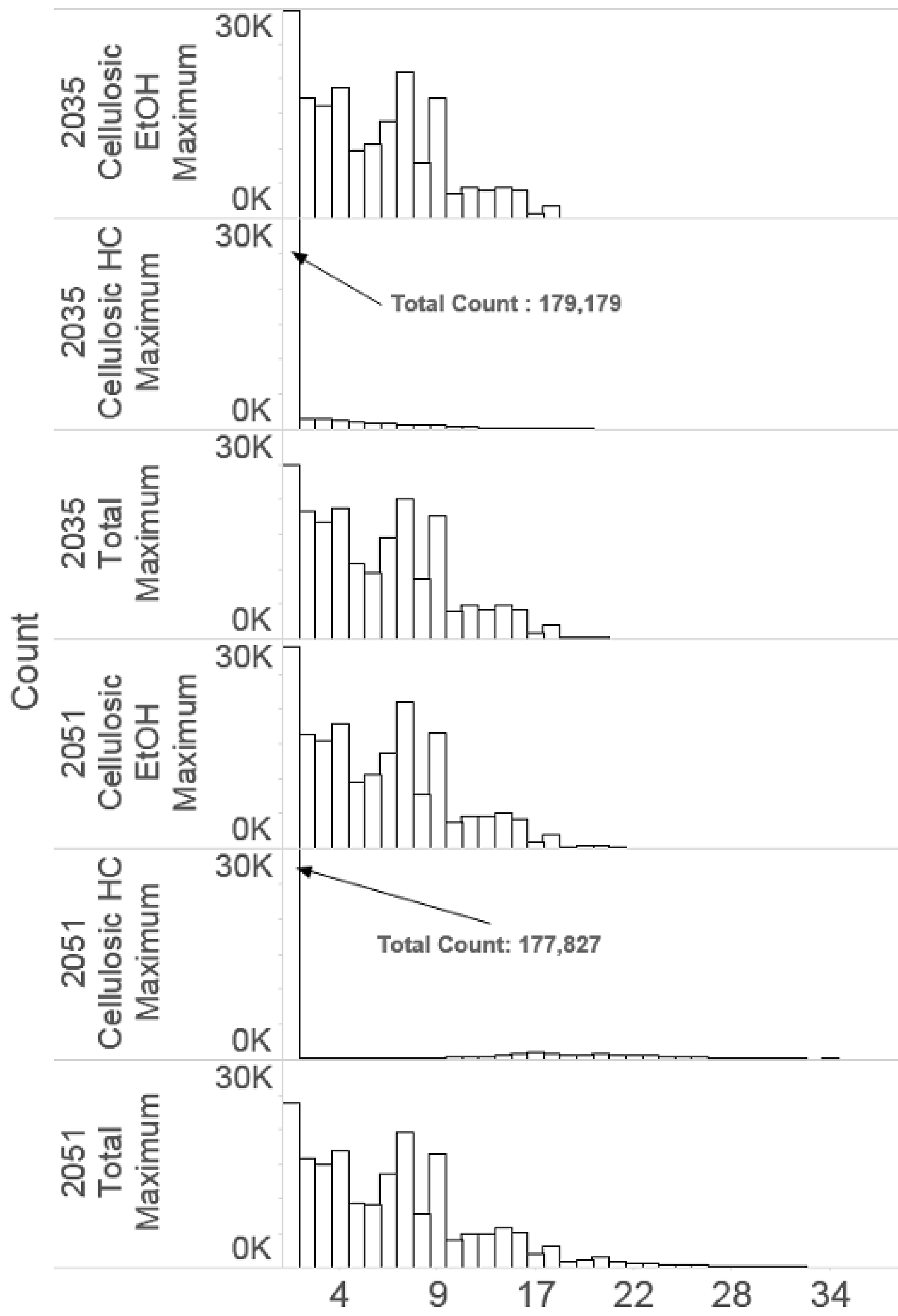

Figure 3: Frequency distribution of fuel production volumes observed among all model simulations. This figure shows the count of simulations ($y$-axis) in each production volume bin (x-axis) for each of the six metrics assessed (rows). The bins show equally-sized ranges of 1 billion gal/yr of fuel production.

### 3.2 Elementary Effects

Results of the elementary effects screening are presented in Figure 4. Of the 140 factors examined, 17 had $\mu^*$ with a standard error greater than 1.2, which suggests that these factors have greater influence than the mean influence of the 140 factors evaluated. These factors are shown in the figure, along with three cellulose to hydrocarbons factors with a standard error greater than 1.0. Our intention was to keep the selection of factors to be used in the VBSA at 20 or fewer to enable a tractable VBSA study design. For example, setting the standard error threshold at 2 resulted in only a few factors being selected, whereas setting the value at 1 resulted in a large number of factors being selected. These threshold values for the standard error of $\mu^*$ were chosen to meet our target.

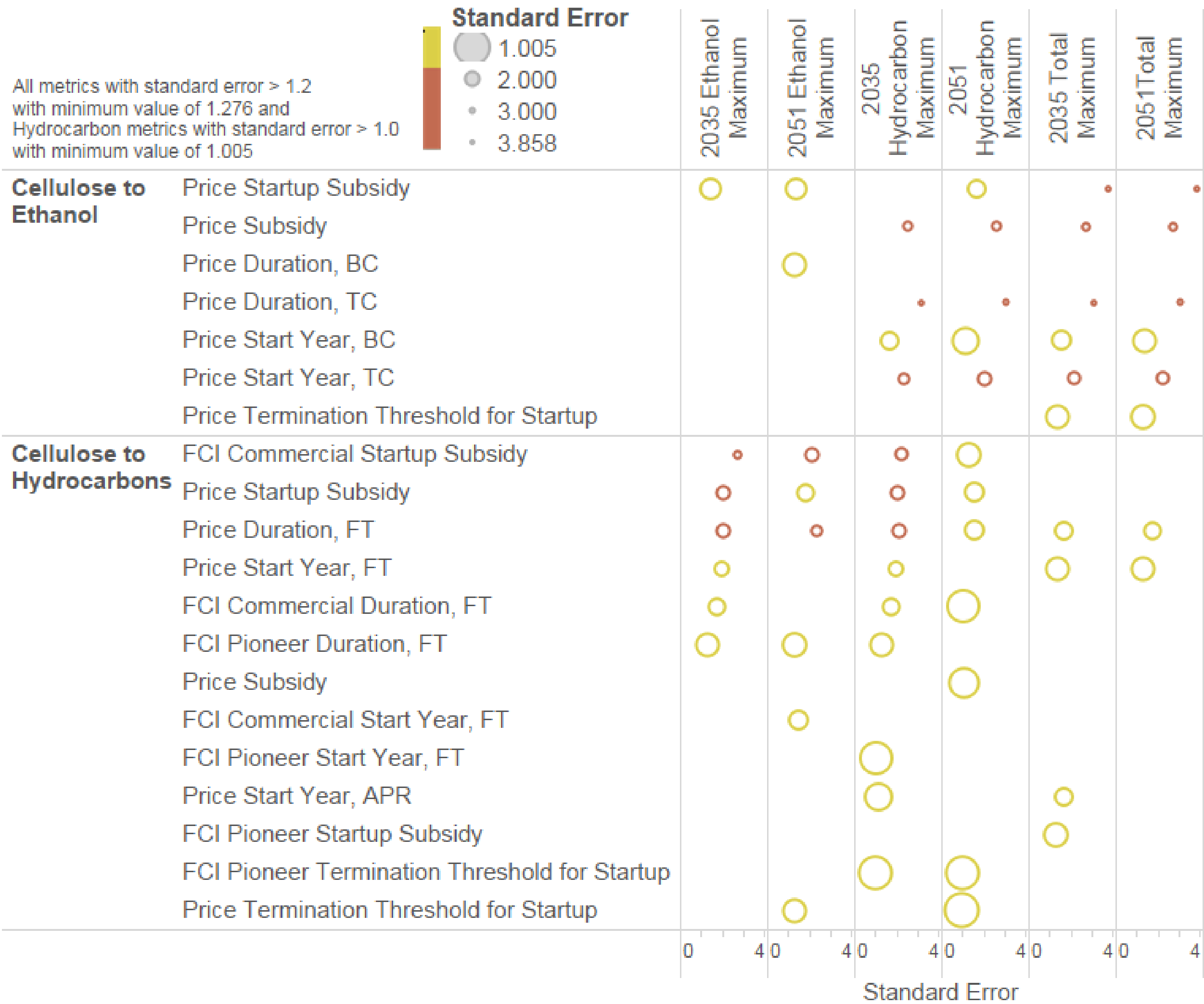

Figure 4: Elementary effects results were used to select 20 potentially influential factors (three controls were also included). The $x$-axis scale is the (unitless) standard error of $\mu^*$. Factors that have larger circles and larger standard error values are more likely to be influential than factors that have smaller circles and lower standard errors.

### 3.3 Sensitivity Analysis

First-order, second-order, and total effects indices were calculated for each of the biofuels production metrics described in Section 2.5. Reported values and confidence intervals for the sensitivity indices are based on bootstrapping.

#### 3.3.1 VBSA First-order Effects

First-order results are presented in Figure 5, which shows the sensitivity indices (S.i) for all six metrics, one in each panel. (S.i in the chart is the same as $S_i$ in the text.) Values are coded by color and symbol to indicate the type of incentive, as described in tables 1, 2, and 3.

Incentives are shown in the same order in each panel. Confidence intervals are indicated by shading. Incentives with large first-order sensitivity index (S.i) values and small confidence intervals indicate greater influence on model results. For first-order indices for maximum total biofuels production prior to 2035 (lower left panel), ethanol-focused incentives that are most influential, in order of magnitude, are start-up price subsidy level, background price subsidy level, and price subsidy duration for biochemical and thermochemical pathways. For maximum total biofuels production prior to 2051 (lower right panel), the most influential factor is the fixed capital investment (FCI) grant levels for biomass-to-hydrocarbons pathways. The ethanol-focused incentives mentioned above are also influential, but their influence is diminished by 2051. (Note that finding an influence on the maximum production before a certain date does not imply a recommendation on the duration or timing of an incentive.)

#### 3.3.2 Interactions: VBSA Second-order Effects

Second-order indices quantify the influence of the pair-wise interactions of factors on the model's output. The set of interactions presented is not exhaustive, and factors might be interacting with other model settings that are not varied in the VBSA. The elementary effects screening criteria allowed certain factors with a $\mu^*$ that is $> 1.2$ standard errors above the mean to be passed on to the VBSA. Strong interactions might be masked if either the interactive pair is not included or only one member of a pair is included in the study design.

Figure 6 shows all combinations of second-order effects for total biofuels production prior to 2035. The blue region highlights interactions among ethanol incentives, the orange region highlights interactions among hydrocarbons incentives, and the green region highlights interactions among ethanol and hydrocarbons incentives.

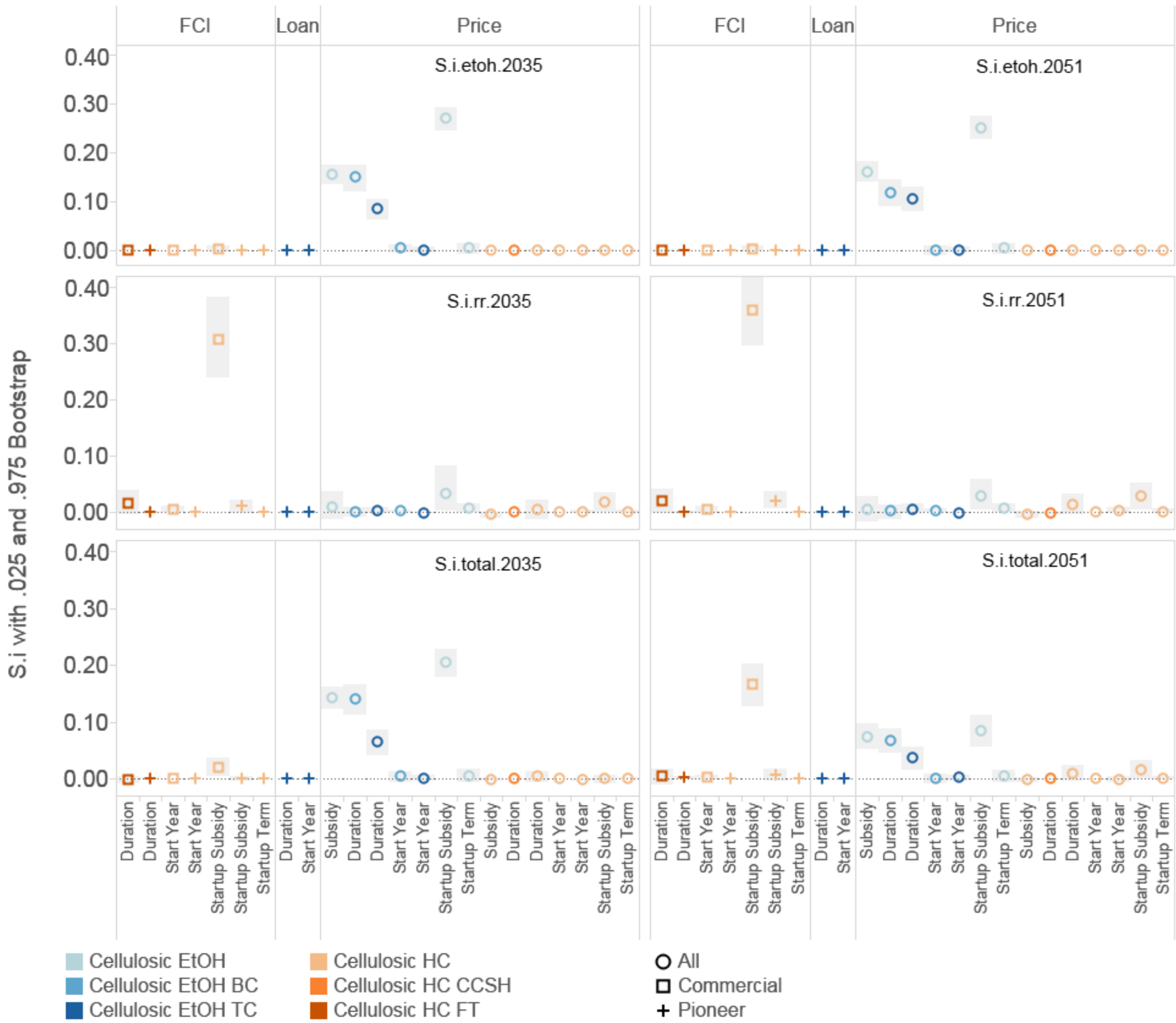

Figure 5: First-order effects (S.i) show relative influence on model variance of each factor alone. S.i in the chart is the same as $S_i$ in the text. Each panel shows results for one biofuel production–year metric. For example, S.i.etoh.2035 refers to the upper left panel as the first-order effects on cellulose-to-ethanol production in 2035 and S.i.rr.2051 refers to the middle right panel as the first-order effects on cellulose-to-hydrocarbons production prior to 2051. The $y$-axis label shows each first-order sensitivity index value and gray shading indicates the bootstrap 95% confidence interval from 0.025–0.975. Each $x$-axis column represents a single factor. The factors are grouped by subsidy type (see labels at top), and they are labeled according to their description and limit (see tables 2 and 3). The color coding shows the feedstock, product, and conversion technology; the different shapes show the scales of production (see Table 1). EtOH = ethanol; BC = biochemical; TC = thermochemical; HC = hydrocarbons; CCSH = catalytic conversion of sugars to hydrocarbons; FT = Fischer Tropsch; rr = refinery-ready, a synonym for hydrocarbons.

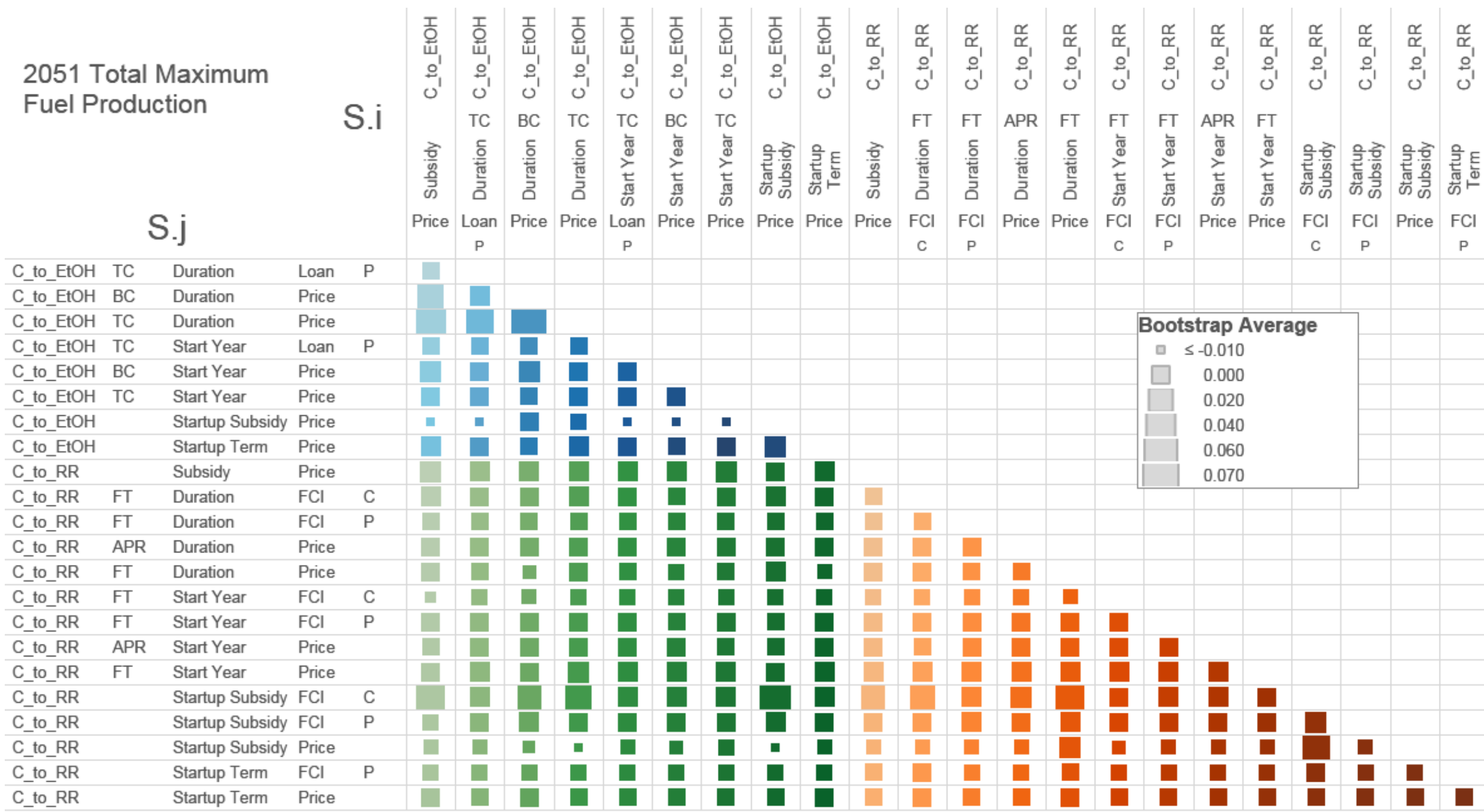

Figure 6: Second-order effects show relative influence on model variance of each pair of factors for total maximum fuel production prior to 2051. The blue region highlights interactions among ethanol incentives; the orange region highlights interactions among hydrocarbons incentives, and the green region highlights interactions among ethanol and hydrocarbons incentives. The size of the bar indicates the size of the second-order sensitivity index, but does not indicate at what confidence level each effect is greater than zero or different in magnitude from others. Colors distinguish each fuel pathway pair. C_to_EtOH = cellulose to ethanol; BC = biochemical; TC = thermochemical; C_to_RR = cellulose to hydrocarbons; APR = catalytic conversion of sugars to hydrocarbons; FT = Fischer Tropsch; P = pioneer commercial; C = full commercial

#### 3.3.3 VBSA Total Effects

Figure 7 shows the total effects for the six metrics. For maximum total biofuels production prior to 2035 (lower left panel), there are three clusters of influential factors. The first consists of ethanol price support duration for biochemical ethanol and the ethanol start-up level. The second cluster comprises ethanol background price subsidy level and ethanol price support duration for the thermochemical pathway. The third cluster includes FCI grant levels for biomass-to-hydrocarbons pathways. For maximum total biofuels production prior to 2051, the most influential factor is the percentage of capital costs that are paid through the FCI for biomass-to-hydrocarbons pathways. A second group of influential factors includes ethanol price support levels and durations as well as price support for the biomass-to-hydrocarbons and FCI duration and level for the Fischer Tropsch pathway.

Figure 8 shows the largest first-order ($\boldsymbol{i}$) and total ($\boldsymbol{T}$) effects in the columns, superimposed on the elementary effects screening. The bootstrapped first-order index range does not include 0 and T applied for the top five total effects indices. The metrics are different in their $\boldsymbol{i}$ and $\boldsymbol{T}$ labeling across the rows, highlighting the value of including multiple metrics. To show changes in the importance of factors during the study period, the $\boldsymbol{T}$ labels in cells lacking circles show the imperfect nature of the elementary effects screening step; these factors would have been excluded by the elementary effects screening based on that metric alone (no circle), but they are among the largest total effects ($\boldsymbol{T}$).

The elementary effects screening reduced the dimensionality of our factor sampling space for the total effects portion of the study. Recognizing the tension between including a number of factors large enough to show the potential interaction effects and small enough for our computing resources and time constraints, our goal was to select approximately 20 factors for the total effects study, based, in part, on [21]; however, we suggest a methodological variant that would (1) perform the elementary effects study; (2) select a subset of the factors for further study that is larger than the 20-factor set size that we used; (3) calculate first-order and total effects of these factors (because first-order and total effects calculations have lower computational requirements); (4) select factors with first-order and total sensitivity indices with 95% confidence intervals that are more than zero and (5) calculate second-order effects (higher computational requirements) for only those factors.

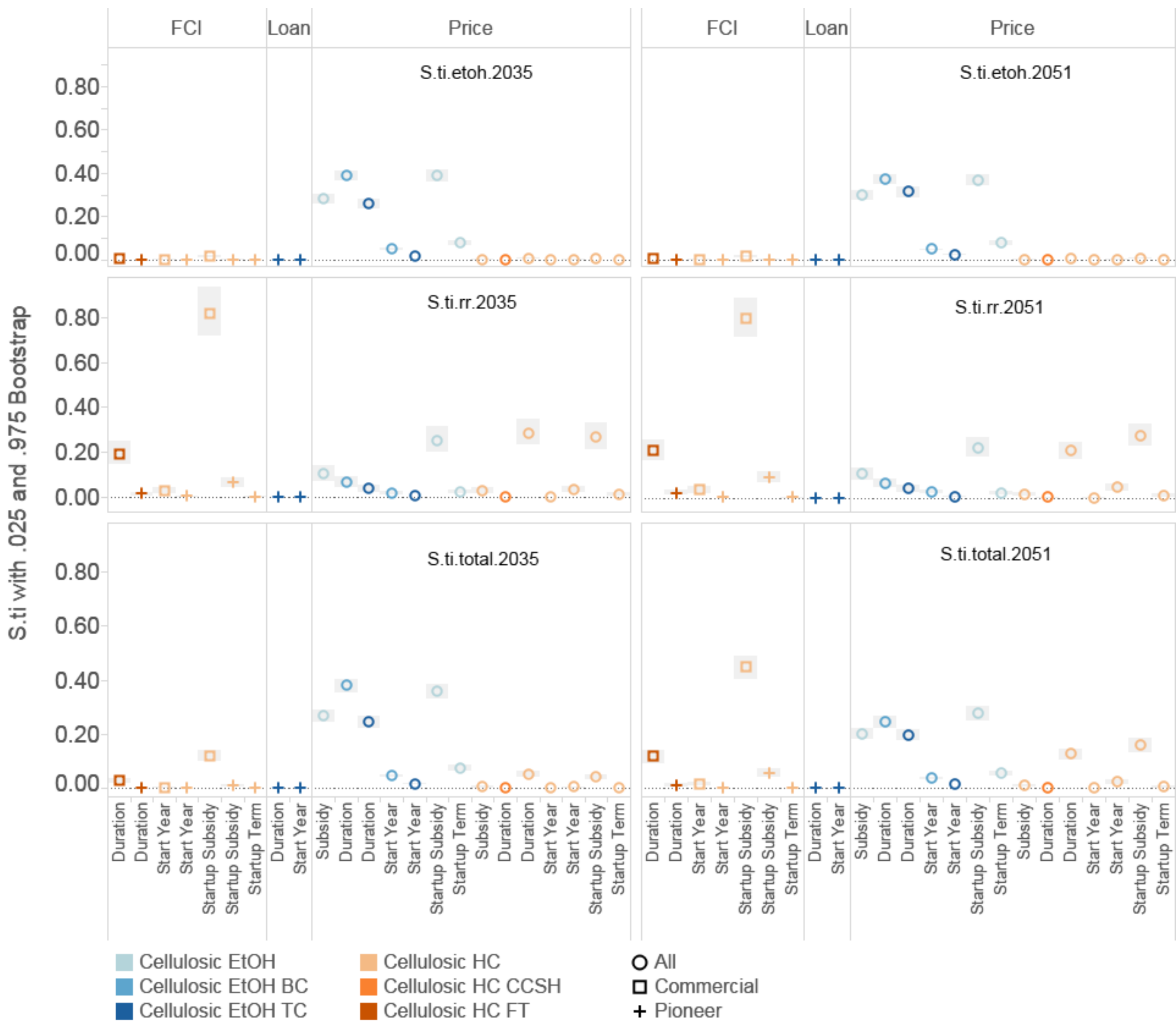

Figure 7: Total effects (S.ti) show relative influence on model variance of each factor alone and in combination with others. S.ti in the chart is the same as $S_{Ti}$ in text. Each panel shows results for one biofuels production–year range metric. For example, S.ti.etoh.2035 refers to the upper-left panel as the total effects on cellulose to ethanol production in 2035 and S.ti.rr.2051 refers to the middle-right panel as the first-order effects on cellulose-to-hydrocarbons production in 2051. The $y$-axis label shows each total effects sensitivity index value and gray shading indicates the bootstrap 95% confidence interval from 0.025–0.975.Each $y$-axis column represents a single factor. The factors are grouped by subsidy type (see labels at top), and labeled according to their description and limit (see tables 2 and 3). The color coding shows the feedstock, product, and conversion technology; the different shapes show the scales of production (see Table 1).

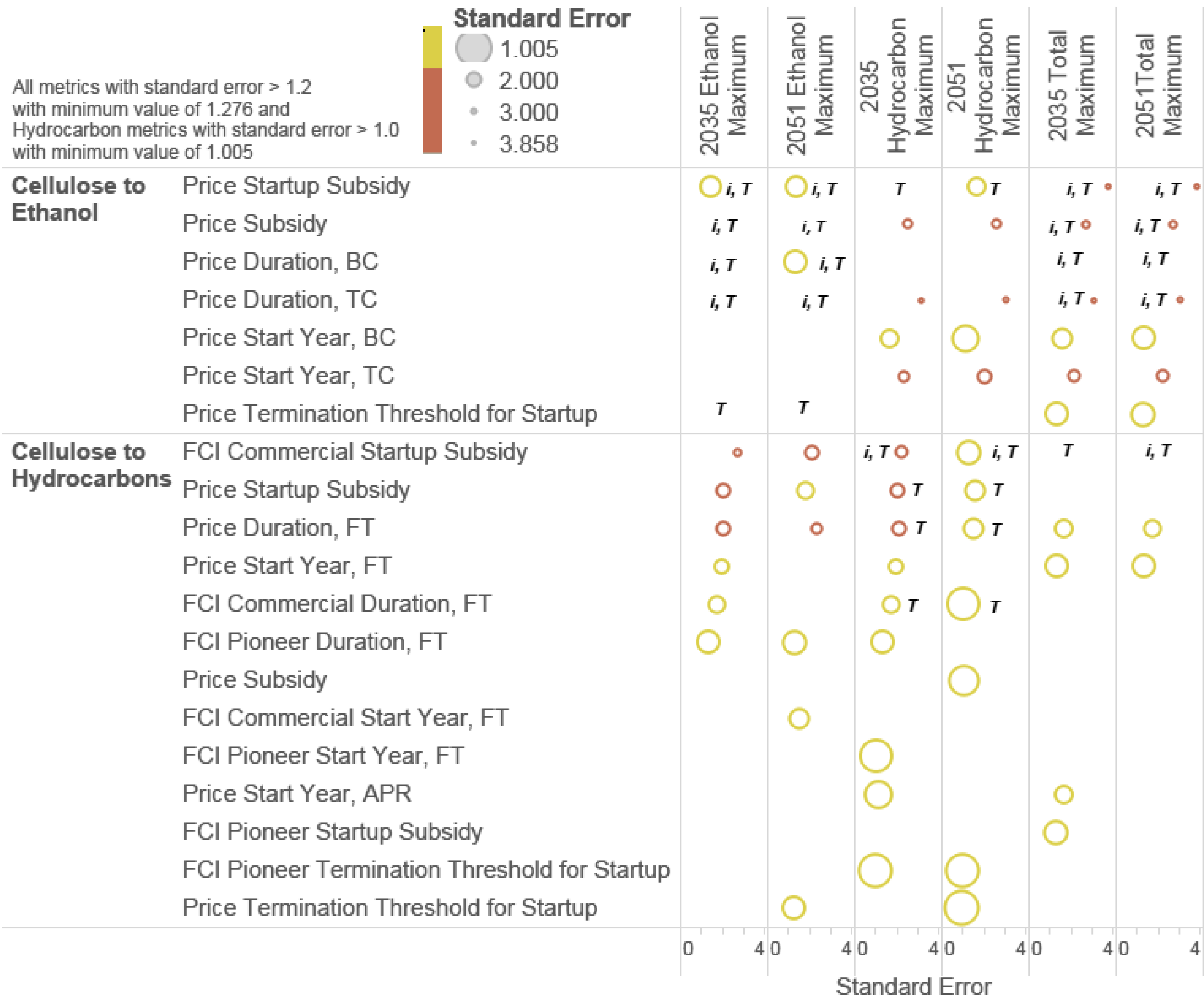

First order and total effects indices overlaid on the factors identified as potentially influential through elementary effects analysis. An "i" indicates that the bootstrapped first-order index range does not include zero for that factor-metric combination. A "T" indicates that the total effects index is among the five highest values for that combination.

Figure 8: First-order and total effects results superimposed on Figure 4. The first-order (***i***) and total (***T***) effects labels apply to the five largest sensitivity index values for each type in each column, if their confidence intervals are more than zero.

### 3.3.4 Sample Size and Uncertainty

For the metrics of maximum biofuels production by both years, the bootstrapped results for the first-order and total effects show some clear distinctions in the estimated ranges. This suggests that the selected sample size of N = 4000 was sufficient for the first-order and the total effects, and it suggests that the rank order of the first-order and total effects are likely to be correct when bootstrap error bars do not overlap and indistinguishable when they do. Although some second-order effects are distinguishable, fewer second-order effects were distinct from zero or distinct from each other, or both, and this could be improved with a larger sample size.

## 4 Discussion

This study presented and demonstrated a novel application and merging of two sensitivity approaches (elementary effects and VBSA) to a large complex model. This approach enables the rigorous quantification of influential model factors that can be be used to inform understanding of the system as a whole. Although these types of studies are rarely performed for models such as the BSM, they should be part of the analyst's regular tool kit. Additionally, we used the BSM as an example in this study, but the approach described is not limited to system dynamics models and could be applied to a wide range of complex computational models.

Sensitivity analysis can improve conceptualization, modeling, and data to understand complex system behavior. As [26] describes: "...system dynamics emphasizes a multifaceted process for testing models, identifying errors, and comparing model assumptions and behavior to data. The process of model testing and improvement is iterative. Discrepancies between mental models, formal models, and data stimulate improvements in each." Complex system dynamics incorporate numerous feedbacks and exhibit nonlinear results. In such models, identifying key system levers and quantifying their strength is not straightforward. VBSA offers a systematic, global approach to assessing system dynamics models because it addresses nonlinear responses and interactive effects. Especially when a large model size makes manual exploration of the input space difficult and time-consuming, the approach can help provide a comprehensive understanding of interactions that drive model behaviors. The quantitative metrics provided by VBSA can help the analyst assess uncertainties and uncover leverage points that might not be obvious in a less comprehensive analysis. The supplementary code, developed in the previous paper and adapted for this study (available in the online supplement), provides a generic and extensible mechanism for performing VBSA. Computational requirements for such analysis, however, might be substantial. Overall, VBSA might be valuable in quantifying and refining understanding of large system dynamics models that have uncertainty in their inputs.